# Signatures of sliding Wigner crystals in bilayer graphene at zero and finite magnetic fields

Anna M. Seiler[1,*], Martin Statz[1], Christian Eckel[1], Isabell Weimer[1], Jonas Pöhls[1], Kenji Watanabe[2], Takashi Taniguchi[3], Fan Zhang[4], R. Thomas Weitz[1,*]

[1]1st Physical Institute, Faculty of Physics, University of Göttingen, Friedrich-Hund-Platz 1, 37077 Göttingen, Germany

[2]Research Center for Functional Materials, National Institute for Materials Science, 1-1 Namiki, Tsukuba 305-0044, Japan

[3]International Center for Materials Nanoarchitectonics, National Institute for Materials Science, Tsukuba, Japan

4Department of Physics, University of Texas at Dallas, Richardson, TX, 75080, USA

*Corresponding authors: seileran@phys.ethz.ch, thomas.weitz@uni-goettingen.de

**Abstract:**

AB-stacked bilayer graphene has emerged as a fascinating yet simple platform for exploring macroscopic quantum phenomena of correlated electrons. Unexpectedly, a phase with negative d$R$/d$T$ has recently been observed when a large electric displacement field is applied and the charge carrier density is tuned to the vicinity of an ultra-low-density van Hove singularity. This phase exhibits features consistent with Wigner crystallization, including a characteristic temperature dependence and non-linear current bias behavior. However, more direct evidence for the emergence of an electron crystal in AB-stacked bilayer graphene at zero magnetic field remains elusive. Here we explore the low-frequency noise consistent with depinning and sliding of a Wigner crystal lattice. The current bias and frequency dependence of these noise spectra align well with findings from previous experimental and theoretical studies on the quantum electron solids. Our results offer transport signatures consistent with Wigner crystallization in AB-stacked bilayer graphene at zero and finite magnetic fields, paving the way for further substantiating an anomalous Hall crystal in its original form.

**Main:**

A recurring theme in condensed matter physics is the exploration of the ground states of a strongly correlated electron system under different conditions [1], particularly when a new system is established in experiment. One candidate state in two dimensions is the so-called Wigner crystal (WC) that minimizes the Coulomb interaction energy of itinerant electrons in the low-density limit [2]. In a WC, due to the dominance of Coulomb repulsion, the itinerant electrons arrange themselves below a melting temperature into a periodic lattice independent of the underlying atomic structure [2]. A central question in the field is how to identify the presence of Wigner crystallization, which is especially critical in situations where direct imaging of an emergent electron crystal is not straightforward [3,4]. An elegant method in this respect is electronic noise measurements: When a sufficiently high external in-plane electric field is applied across the sample, the electron lattice pinned by charge impurities can become mobile. This depinning triggers a sliding behavior, where the Wigner crystal phase experiences gradual, collective motion in response to the external perturbation [5–7]. This motion gives rise to an oscillatory current determined by the sliding velocity and the lattice constant of the emergent WC, and this unique behavior is often detectable as noise in DC transport measurements [5–7]. Therefore, detecting an AC current with a frequency that is proportional to the applied bias current serves as a clear signature of sliding WC's [8–14].

To date, the investigation of noise in charge ordered states has mainly focused on quasi-one or two dimensional bulk materials, such as WC phases at large magnetic fields [5,11,14–19] and incommensurate CDW phases in transition metal dichalcogenides [8,9,20]. However, recent advancements have expanded the scope of exploring correlated electronic ordering in even simpler systems that neither require a magnetic field to quench the kinetic energy nor a moiré superlattice to trap the emergent electron crystal. Specifically, the naturally occurring rhombohedral-stacked multilayer graphene systems have emerged as an outstanding platform for studying strongly correlated electrons [4,21–26]. Not only is this evident when a large magnetic field is applied [4], but this has also been identified even at zero magnetic field when the low charge carrier density is tuned to the vicinity of van Hove singularities [21–23]. While magnetic field-induced WC's have been observed using high-resolution scanning tunneling microscopy (STM) in AB-stacked bilayer graphene (BLG) [4], compelling evidence for Wigner crystal states at zero magnetic field remains elusive in graphene systems, although bias spectroscopy and temperature dependent resistance measurements have provided first hints towards correlated phases consistent with such ordering in AB-stacked BLG [21,22] and rhombohedral pentalayer graphene [23]. To provide further evidence for Wigner crystallization in BLG with transport measurements, one smoking-gun experiment is to investigate the low-frequency noise generated by depinning and sliding of emergent electron lattices at zero magnetic field near the isospin Stoner phase regime [21,22,27] and at large magnetic fields near the fractional quantum Hall regime [4].

The BLG flake investigated in this study is encapsulated in hexagonal boron nitride (hBN) and equipped with graphite top and bottom gates and two-terminal graphite contacts (the device and its fabrication were detailed in Ref. [21]). This configuration enables to continuously tune the vertical electric displacement field $D$ and the charge carrier density $n$ (see Methods). All measurements, unless stated otherwise, were conducted in a dilution refrigerator at a base temperature of 10 mK using a standard lock-in technique (see Methods).

To establish the measurement technique in our BLG flake and validate their capability in identifying WC phases, we first apply them to the fractional quantum Hall regime at large magnetic fields (Fig. 1a,b) in which a recent high-resolution STM study has unambiguously identified a WC phase [4]. While the details of the measurements are described below, at the out-of-plane magnetic field $B_\perp$ = 14 T and in the vicinity of $n$ = -0.6 x $10^{11}$ cm$^{-2}$, a three order of magnitude increase of the spectral current noise density can be detected at low DC current of 0.5 nA (Fig. 1c). This provides compelling evidence for the presence and the depinning of the magnetic field induced WC near the fractional filling factor -1/5, consistent with the reported findings based on STM measurements [4].

Hereafter we shall examine the zero magnetic field regime with the same technique, where an unambiguous proof of a WC phase is still missing even though initial indications consistent with electron crystallization have emerged from bias spectroscopy and temperature-dependent resistance measurements [21,22]. We first reproduce the previous observations of the low-temperature phase diagram close to the WC phase to be studied. Fig. 2a shows the measured two-terminal conductance map $G$ in arbitrary units (a.u.) at $D$ = -0.7 Vnm$^{-1}$ as a function of $n$ and $B_\perp$ measured with an applied AC current of 1 nA. At low $B_\perp$ and hole doping, a complex phase diagram emerges because of the interplay between electron-electron interaction and trigonal warping induced van Hove singularities [21,28–30], in the vicinity of which a variety of new correlated phases appear. Of special relevance for this work is an emergent phase in the vicinity of the fully isospin polarized van Hove singularity (phase II in Fig. 2a [21]) that manifests strong non-linearities with applied bias current and an insulating temperature dependence [21]. Both features are consistent with a WC phase. More attractively, this phase in the Landau fan diagram not only emanates from a range of finite densities at $B$ = 0 but also follows the Středa formula with a quantized slope at finite $B_\perp$ fields [21]. Such topological phases, dubbed Wigner-Hall crystals [21] or anomalous Hall crystals [31–33], have sparked strong interest in the field. Here we aim to substantiate the WC nature of phase II.

A time-trace of the conductance within the density region of phase II is shown in Fig. 2b and Fig. 3c. Pronounced fluctuations in the conductance can be observed at an applied low-frequency (78 Hz) AC bias current of 1 nA, consistent with the sliding of a potential WC, as shown by the orange linecut in Fig. 2b. Crucially, these fluctuations manifest only within the density range associated with phase II (see the purple, black and green linecuts in Fig. 2b taken outside phase II) and become less pronounced in regions where phase II is destabilized by decreasing $D$ (Extended Data Fig. 1) or increasing temperature (Extended Data Fig. 2). The fluctuations are negligible below AC bias currents of 100 pA (Extended Data Fig. 3), giving a first hint towards possible WC depinning and sliding behavior at larger AC biases to be investigated below. In the density region of the resistive phase III, the conductance fluctuations are much weaker (Fig. 2b, Extended Data Fig. 1). This phase also exhibits a much weaker dependence on temperature and current bias compared to phase II. Phase III is likely a potential WC phase with a much weaker pinning potential, and the depinning occurs at a much lower bias current [21]. Nevertheless, we shall focus on the further investigations of phase II.

To gain insights on the sliding behavior of the potential WC phase, we investigate the influence of an applied bias current on the resistance fluctuations (Fig. 3 and 4). We apply DC currents ($I_{DC}$) on top of a constant 78 Hz AC current ($I_{AC}$) of 100 pA at $D$ = -0.7 Vnm$^{-1}$ and monitor the conductance (Fig. 3a, b) and its changes in the normalized resistance over time $\frac{dR}{dt} \times \frac{1}{R_0}$, with $R_0$ taken at time $t$ = 0 s (Fig. 3c). There are no resistance fluctuations observed at low currents, indicative of a stable region where no depinning takes place. Despite that the conductance remains nearly constant up to $I_{DC}$ = 3.5 nA within the density range of phase II (Fig. 3a, b), significant temporal fluctuations become evidenced at $I_{DC} \geq 1$ nA. Importantly, these fluctuations span the entire density range of phase II. This behavior is consistent with the depinning and sliding of a WC at applied DC currents of 1 nA to 5 nA. Above 4 nA, $\frac{dG}{dI_{DC}}$ exhibits a jump, which may indicate a phase slip induced by the large current. Notably, resistance fluctuations are also present at higher DC currents (~10 nA) near the phase boundary between phases II and phase I, and they can be attributed to the abrupt change of conductance (Fig. 3c).

The resistance fluctuations become increasingly pronounced when a small in-plane magnetic field $B_{\parallel}$ is applied, as exhibited in Fig. 4a-c. A small $B_{\parallel}$ can strengthen phase II [21] and consequently shift the threshold current to a higher value of $I_{DC}$. Intriguingly, resistance fluctuations are observed only at $I_{DC} \approx 5$ nA at $B_{\parallel}$ = 200 mT (Fig. 4c), suggesting the onset of depinning and the onset of possible phase slip occur at nearly the same $I_{DC}$. More intriguingly at $B_\parallel$ > 300 mT, no prominent resistance fluctuations are present and it appears that the possible phase slips could occur at even lower $I_{DC}$ values than the onset of depinning. These features deserve more elaborated examinations in the future.

In case of depinning and sliding WC's, the periodic WC lattice structure is anticipated to induce a periodic modulation of the current with a characteristic sliding frequency known as the washboard frequency $f_0$, which is determined by the time-averaged velocity $v$ of the moving WC and the lattice constant $a_0$ of the WC: $f_0 = \frac{v}{a_0}$ [5,9,11]. $a_0$ depends on $n$ via $a_0 = \sqrt{\frac{2}{\sqrt{3}n}}$, assuming a triangular lattice for the Wigner crystal [4]. $v$ is also given by $v = \frac{j}{en}$ with current density $j$ and electron charge $e$. Finally, the characteristic frequency at which the charge carriers' slide is given by $f_0 = \frac{v}{a_0} = \sqrt{\frac{\sqrt{3}}{2}} \frac{J}{e\sqrt{n}} = \sqrt{\frac{\sqrt{3}}{2}} \frac{I}{ew\sqrt{|n|}} \approx 5 \times 10^7$ Hz in our sample with sample width $w$ = 3 μm, $I$ = 1 nA, $|n|$ = 1.3 x $10^{11}$ cm$^{-2}$ and $a_0$ = 29.8 nm.

To investigate the periodic modulation of the current and to experimentally determine $f_0$, we analyze the normalized spectral noise density $S$, i.e., a measure of the normalized variance of the measured voltage (see Methods). We employ an AC-DC interference technique, where a 100 pA AC signal with a tunable frequency $f$ is superimposed on the DC driving current $I_{DC}$. This approach serves as a sensitive probe of the washboard frequency, enabling the detection of DC-dependent resonances [14]. It is noteworthy that the measured frequency noise is usually much smaller (approximately three to four orders of magnitude) than the expected $f_0$ which is explained with the motion of Wigner crystal domains (the systems is therefore sometimes called a Wigner solid rather than a Wigner crystal) and its interaction with the disorder potential rather than the motion of an entire, defect free crystal [9,11]. Therefore, although we expect $f_0 \approx$ 10 MHz at $n$ = -1.3 x $10^{11}$ cm$^{-2}$ and I = 1 nA, we restricted the frequency range to $f$ < 1370 Hz (electrical filters integrated into our cryostat cut out high frequencies), and within this range $G$ remains nearly constant (Extended Data Fig. 4).

In the absence of sliding, i.e., at $I_{DC}$ = 0 nA and $I_{DC}$ > 5 nA (Fig. 5a), and outside the density range of phase II (Extended Data Fig. 4c), $S$ is predominantly characterized by a $1/f$ or "flicker" noise spectrum, as shown by the linecuts for $I_{DC}$ = 0 nA, $I_{DC}$ = 20 nA and $I_{DC}$ = 80 nA in Fig. 5a. This type of noise spectrum is commonly observed across a frequency range spanning from a few Hz to tens of kHz in various materials and arises from different fluctuation processes, such as mobility fluctuations caused by scattering centers within the substrate or sample [34]. Note that there are additional peaks around 50 Hz and its higher harmonics, which can be attributed to electrical power distribution noise (the standard mains frequency in Germany is 50 Hz).

Deviations from the $1/f$ noise spectrum appear in phase II within its specific ranges of $I_{DC}$ and $n$, where fluctuations in $R$ over time are evident (Fig. 3, 4, 5). In this regime, $S$ exhibits a significant increase by orders of magnitude (Fig. 5a). To separate the enhanced noise spectrum from the underlying $1/f$ background noise, Fig. 5b displays the normalized spectral noise density, i.e., $S$ divided by $S$ ($I_{DC}$ = 80 nA), as a function of $f$ for different $I_{DC}$. The appearance of noise bulges, each centered around a characteristic frequency $f_0$, and their evolution with increasing $I_{DC}$ are clearly visible. This effect is further amplified (Fig. 5c) when small $B_{\parallel}$ of 20 mT are applied to strengthen the phase II [21].

The noise bulges exhibit an inhomogeneous broadening, likely due to the presence of competing processes with different time constants [9]. To determine $f_0$, we apply polynomial fits to the different spectra (note that Lorentzian noise spectra would be expected without the presence of inhomogeneous broadening [9]). Fig. 5d illustrates the trend of $f_0$ as a function of $I_{DC}$ for $B_{\parallel}$ = 0 T and $B_{\parallel}$ = 20 mT. Consistent with the depinning and sliding behavior of WC, $f_0$ increases linearly with $I_{DC}$ with and without the applied $B_{\parallel}$, providing another compelling transport signature for the Wigner crystallization in phase II. These measured $f_0$ values are consistent with that of a WC in GaAs quantum wells [11]. Similar to this well-established case [11], if the time-averaged velocity is interpreted as that of the moving WC, the estimated $f_0$ is four orders of magnitude larger. This mismatch, similarly present in GaAs quantum wells and AB-stacked BLG, calls for future theoretical inspection.

Finally it is noteworthy that we identified similar signatures of WC depinning and sliding also in electron-doped AB-stacked BLG [22] (Extended Data Fig. 5) using the same bias-dependent noise measurements applied here to the hole-doped regime. Additionally, we note that in other AB-stacked BLG experiments no clear signs consistent with WC have been found [35]. In a similar region of the phase diagram, but at higher temperatures and over a broader density range, the formation of magnetic domains has been observed [35]. While it remains to be investigated whether such domains could produce the type of characteristic noise detected in our experiments, it seems unlikely since at least in noise studies of spin transfer torque random telegraph noise with a 1/f spectrum has been observed [35–38] – at odds with the low-frequency noise observed for depinning of WC domains. Potentially the ground state of AB-stacked BLG depends on the strength of the exchange interaction, which is screened by nearby metallic gates, as well as subtle variations in sample quality in this region of the phase diagram.

In conclusion, the analysis of the noise spectra aligns well with previous experimental and theoretical studies on the depinning and sliding WC [6–18,20,34], consistent with the interpretation that the low-density, insulating phase II observed in AB-stacked BLG at zero magnetic field, indeed, originates from Wigner crystallization. No matter the out-of-plane or the in-plane magnetic field increases the depinning onset, however, shifts to higher currents and eventually becomes unobservable at a critical magnetic field (Fig. 4 and Extended Data Fig. 6). A particular intriguing feature of phase II (Fig. 2-5) is its dependence on the out-of-plane magnetic field, which exhibits consistency with a low-density quantum anomalous Hall state with a Chern number of 2 [21]. This underscores the need for alternative methods to verify or dispute the topologically non-trivial nature of phase II in AB-stacked BLG. It is expected that WC phases may also arise in thicker graphene systems with rhombohedral stacking because of the electric gate tunability of the ultra-flat bands near their band gaps at charge neutrality.

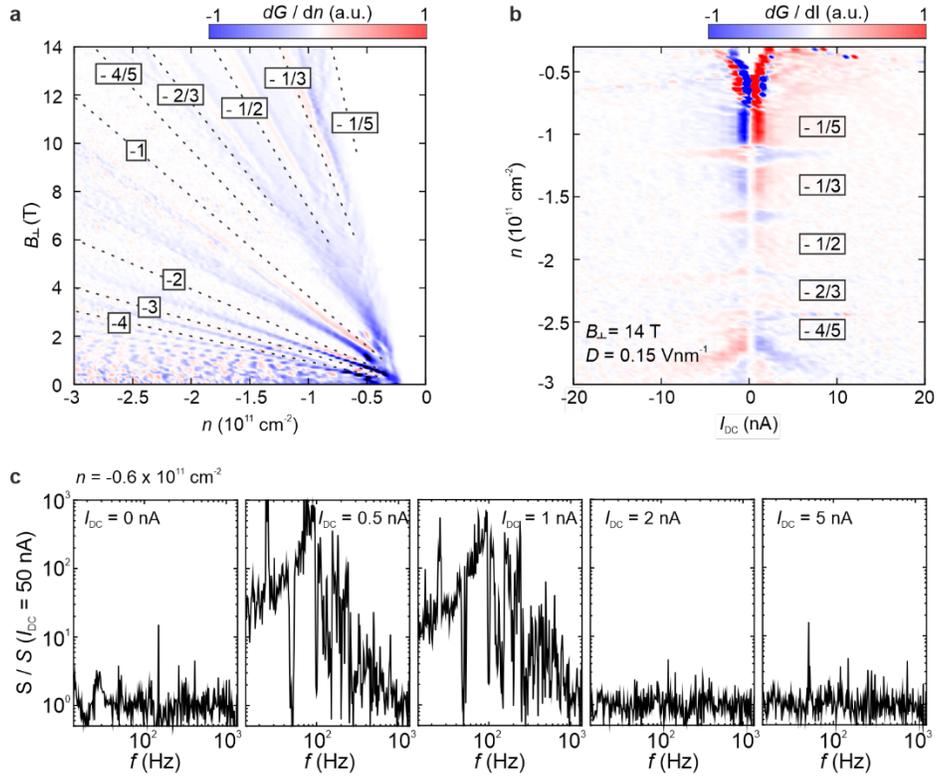

**Fig. 1. Noise in the fractional quantum Hall regime of AB-stacked bilayer graphene. (a)** Derivative of the conductance (d$G$/d$n$) as a function of the charge carrier density $n$ and the out-of-plane magnetic field $B_\perp$ at an electric displacement field of $D$ = 0.15 Vnm$^{-1}$. Integer and fractional quantum Hall states (QHS) emerge at finite $B_\perp$. Fractional QHS with filling factors $v < 1$ and integer QHS with $v \leq 4$ are traced by dashed lines. **(b)** d$G$/d$I$ as a function of the DC bias current $I_{DC}$ and the charge carrier density $n$. Insulating behavior (decreasing conductance with decreasing current) is observed within the QHS [39] and close to the bandgap where a previous STM experiment revealed Wigner crystallization [4]. **(c)** Spectral noise density $S$ normalized with respect to $S$ ($I_{DC}$ = 50 nA) as a function of the applied AC frequency $f$ ($I_{AC}$ = 100 pA) measured at different $I_{DC}$, $n$ = -0.6 x 10$^{11}$ cm$^{-2}$, $B_\perp$ = 14 T and $D$ = 0.15 Vnm$^{-1}$. Frequency-dependent noise bulges appear for $I_{DC}$ = 0.5 nA to $I_{DC}$ = 1 nA.

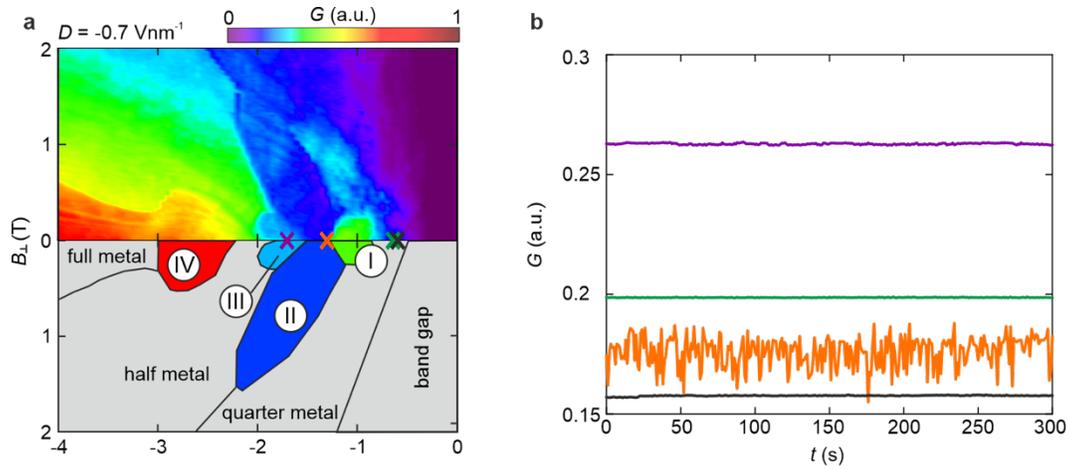

**Fig. 2. Observation of noise in AB-stacked bilayer graphene at zero magnetic field. (a)** Conductance $G$ as a function of charge carrier density $n$ and out-of-plane magnetic field $B_\perp$ at an electric displacement field at $D$ = -0.7 Vnm$^{-1}$ and an AC current (78 Hz) of 1 nA ($I_{DC}$ = 0 nA). The various phases are schematically illustrated and labeled in the mirror image according to reference [21]. **(b)** Linecuts of the conductance as a function of time $t$ at zero magnetic field for the selected densities indicated by colored crosses in **(a)**: $n$ = -1.7 x 10$^{11}$ cm$^{-2}$ (purple, phase III), $n$ = -1.3 x 10$^{11}$ cm$^{-2}$ (orange, phase II), $n$ = -0.65 x 10$^{11}$ cm$^{-2}$ (green, quarter metal), and $n$ = -0.6 x 10$^{11}$ cm$^{-2}$ (black, quarter metal).

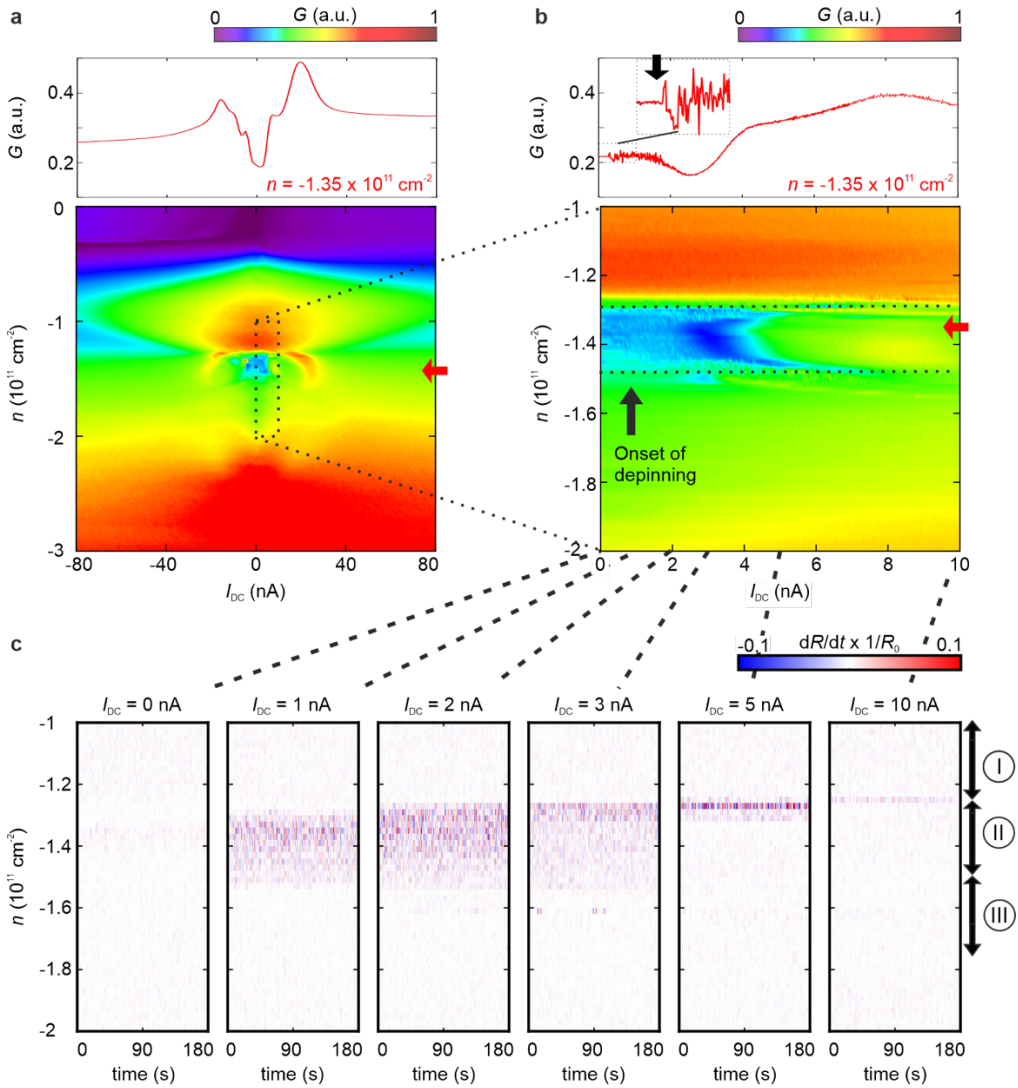

**Fig. 3. Depinning as a function of the in-plane electric field. (a)** Conductance as a function of an applied DC bias current $I_{DC}$ and the charge carrier density $n$ at an electric displacement field $D$ = -0.7 Vnm$^{-1}$ and an AC current $I_{AC}$ = 100 pA. **(b)** Zoom-in of the density region of phase II in **(a)** (marked by dashed lines). Linecuts taken at $n$ = -1.35 x 10$^{11}$ cm$^{-2}$ (marked by red arrows) are shown in the top panel. Noise starts to appear at $I_{DC}$ ≈ 200 pA. The onset of conductance fluctuations is marked with a black arrow in the zoom-in. **(c)** Derivative of the normalized resistance over time $\frac{dR}{dt} \times \frac{1}{R_0}$ (the resistance $R$ is normalized with respect to $R_0$ taken at time $t$ = 0 s) at different $I_{DC}$. The density region of phases I - III (determined at $I_{DC}$ = 0 nA) is indicated by arrows.

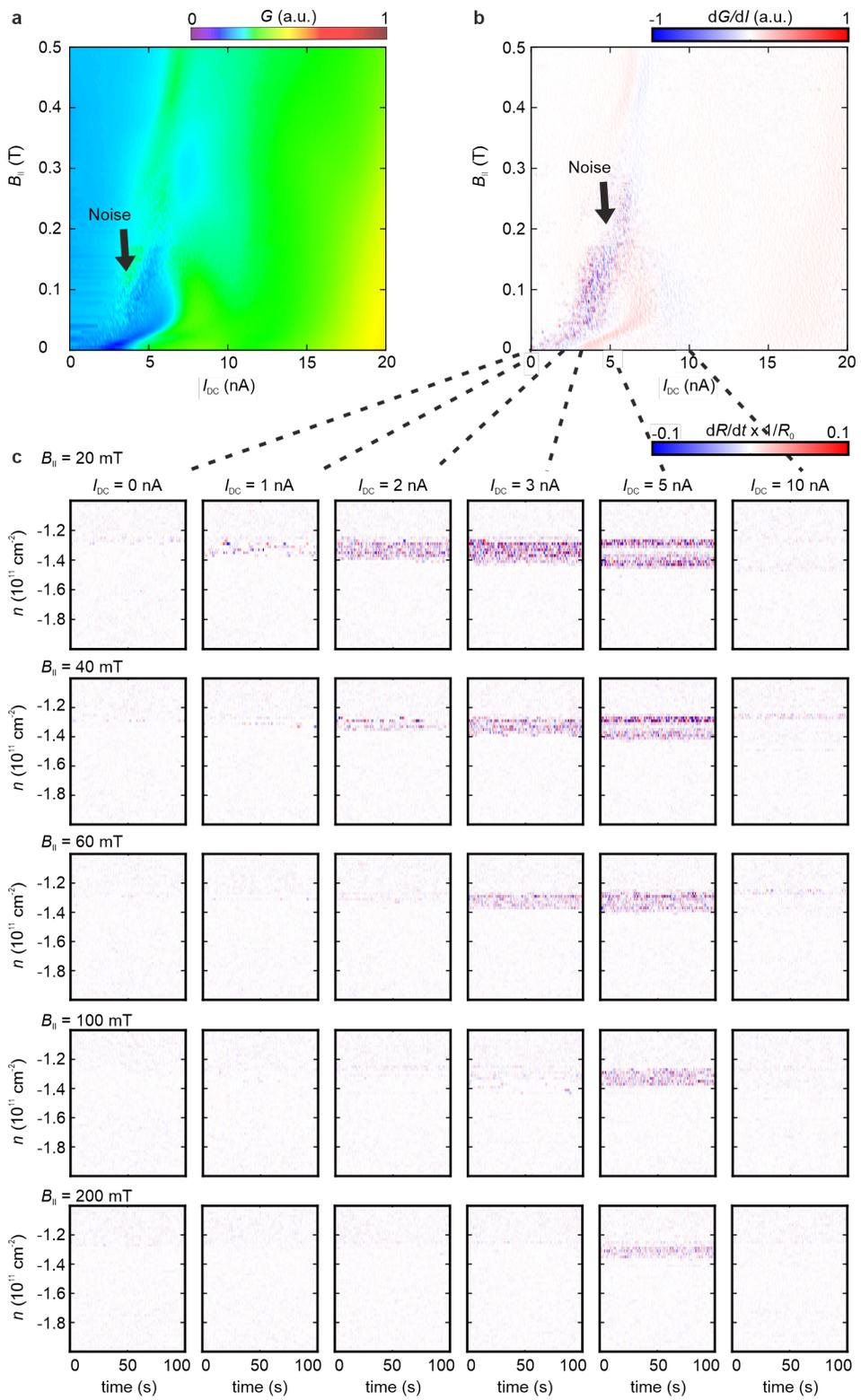

**Fig. 4. Noise as a function of the in-plane magnetic field. (a,b)** Conductance ($G$) **(a)** and derivative of the conductance with respect to the applied DC current $I_{DC}$ ($dG/dI$) **(b)** as a function of $I_{DC}$ and an in-plane magnetic field $B_{\parallel}$ taken at a charge carrier density $n$ = -1.3 x $10^{11}$ cm$^{-2}$ and an electric displacement field $D$ = -0.7 Vnm$^{-1}$. Current fluctuations, labeled as noise, emerge at $B_{\parallel}$ < 0.3 T and $I_{DC}$ < 5 nA before the Wigner crystal is fully depinned at lager $I_{DC}$. **(c)** Derivative of the normalized resistance over time $\frac{dR}{dt} \times \frac{1}{R_0}$ as a function of the time $t$ and $n$ at different $I_{DC}$ and $B_{\parallel}$.

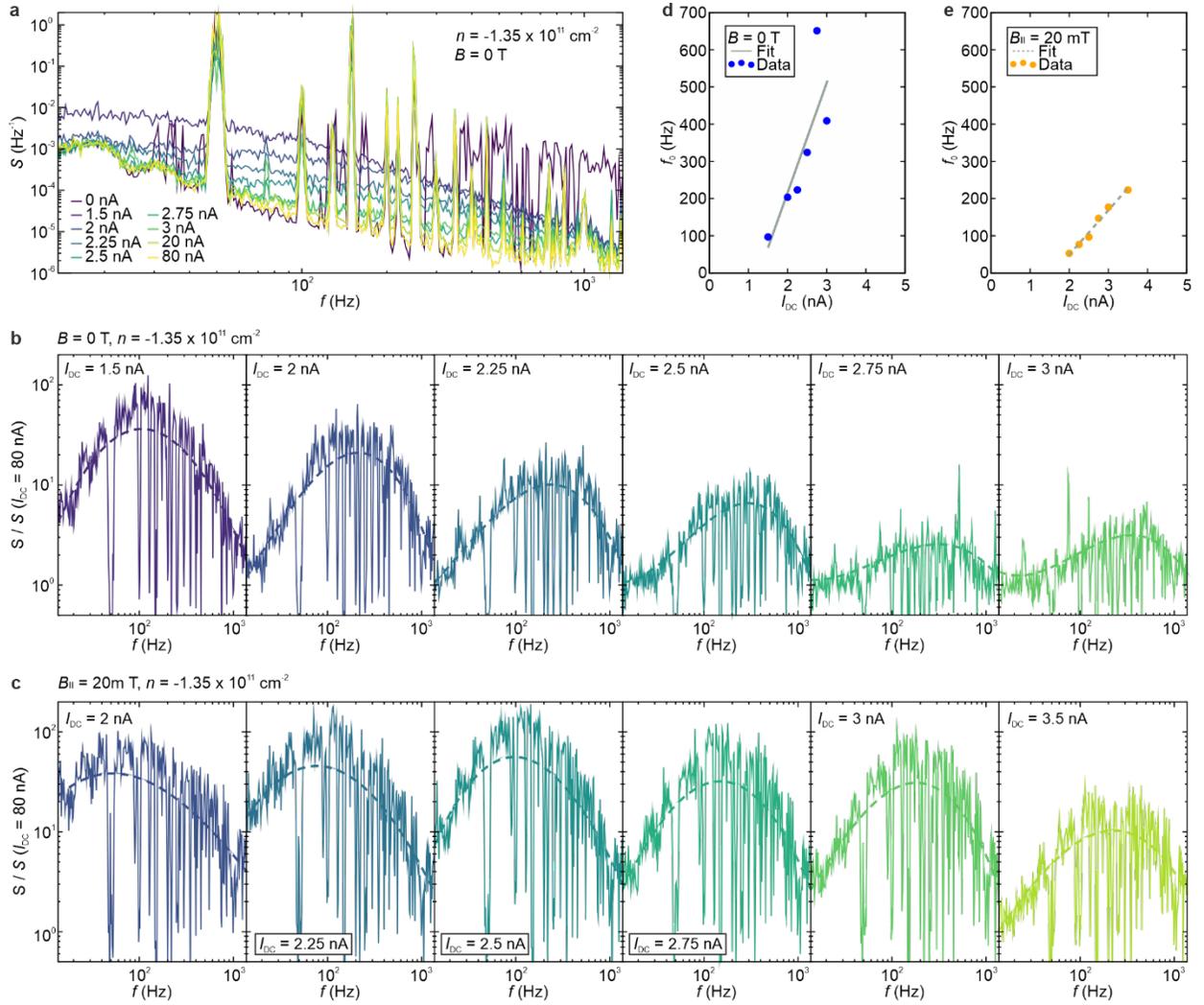

**Fig. 5. Frequency dependence of the current fluctuations. (a)** Normalized spectral noise density $S$ as a function of the applied frequency $f$ of the AC current $I_{AC}$ at different $I_{DC}$ at zero magnetic field $B$, an electric displacement field $D$ = -0.7 Vnm$^{-1}$ and a charge carrier density of $n$ = -1.35 x 10$^{11}$ cm$^{-2}$ within phase II. **(b, c)** Spectral noise density normalized relative to the background spectral noise density measured at $I_{DC}$ = 80 nA ( $S$ / $S$ ($I_{DC}$ = 80 nA) ) as a function of $f$ at different $I_{DC}$ and an in-plane magnetic field B$_{\parallel}$ of 0 T **(b)** and 20 mT **(c)**. The noise bulges are fitted with polynomial functions represented by dashes lines. **(d,e)** Dependence of the washboard frequency $f_0$ at $B_{\parallel}$ = 0 T **(d)** and $B_{\parallel}$ = 20 mT **(e)** extracted from **(b)** and **(c)** as a function of $I_{DC}$. Linear fits are shown in grey.

## Methods:

**Electrical measurements**

All electrical measurements were conducted using a standard lock-in technique. An AC reference signal with a frequency of 78 Hz (unless otherwise specified) was generated by a lock-in amplifier (SR865, Stanford Research Systems) and converted into a small AC current signal of 100 pA - 1nA via a high resistor. The voltage drop across the contacts was then measured using a second lock-in amplifier. An additional DC bias current was introduced using a DC source-measure unit (SourceMeter 2450, Keithley) and a resistor This DC bias was modulated onto the AC reference signal via a home build transformer.
To prevent grounding loops, the measurement units were isolated from standard power lines using isolating transformers.

We can independently tune the charge carrier density *n* and the electric displacement field *D* using graphite top and bottom gates. The charge carrier density *n* is defined as $n = \varepsilon_0\, \varepsilon_{hBN}\, (V_t/d_t + V_b/d_b)/e$, where $V_t$ and $V_b$ are the gate voltages applied to the top and bottom gates, respectively, $d_t$ and $d_b$ are the thicknesses of the upper and lower hBN flakes serving as dielectrics., *e* is the charge of an electron, $\epsilon_{hBN}$ is the dielectric constant of hBN, and $\epsilon_0$ is the vacuum permittivity. The vertical electric displacement field *D* is defined as $D = \varepsilon_{hBN}\, (V_t/d_t - V_b/d_b)/2$.

The normalized spectral noise density *S* is defined as $S = v^2/(ENBW \times V_{Contacts}^2)$, where $v$ represents the variance determined from the lock-in amplifier, ENBW = 0.78 Hz is the equivalent noise bandwidth, and $V_{Contacts}$ is the voltage drop between the contacts. The variance $v$ was determined using an integration time of 20 s, corresponding to at least 200 data points for each chosen frequency.

**Data availability**

Relevant data supporting the key findings of this study are available within the article. All raw data generated during the current study are available from the corresponding authors upon request.

**Acknowledgements**

We acknowledge illuminating discussions with E.Y. Andrei, A. Gosh, A.F. Young and D. Steil. R.T.W. and A.M.S. acknowledge funding from the Deutsche Forschungsgemeinschaft (DFG, German Research Foundation) under the SFB 1073 project B10. R.T.W. acknowledges funding from the DFG SPP 2244. K.W. and T.T. acknowledge support from the JSPS KAKENHI (Grant Numbers 21H05233 and 23H02052) and World Premier International Research Center Initiative (WPI), MEXT, Japan. F.Z. acknowledges supports from US National Science Foundation under grants DMR-1945351, DMR-2324033, and DMR-2414726.

**Author contributions statement**

A.M.S. fabricated the device. A.M.S. and M.S. conducted the measurements with assistance from C.E., I.W., and J.P.. A.M.S. performed the data analysis with help from M.S., C.E., I.W., and J.P.. K.W. and T.T. grew the hexagonal boron nitride crystals. All authors discussed and interpreted the data. R.T.W. supervised the experiments and the analysis. A.M.S., F.Z. and R.T.W. prepared the manuscript with input from M.S., C.E., I.W. and J.P., and feedback from all authors.

**Competing interests**

Authors declare no competing interests.

**Corresponding Authors**

R. Thomas Weitz (thomas.weitz@uni-goettingen.de) and Anna M. Seiler (seileran@phys.ethz.ch).

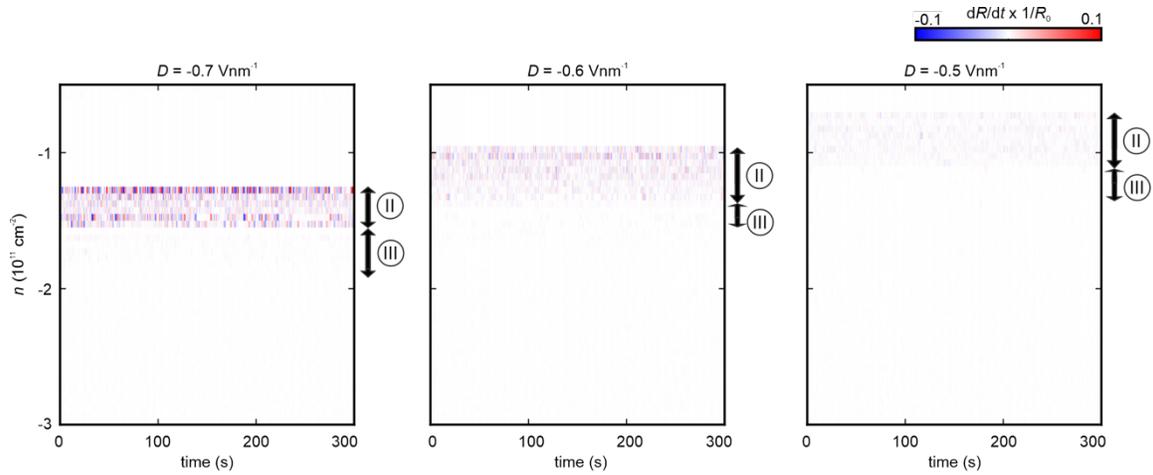

**Extended Data Fig. 1. Noise at different electric displacement fields.** Derivative of the normalized resistance over time $\frac{dR}{dt} \times \frac{1}{R_0}$ as a function of the charge carrier density *n* at different electric displacement field *D*, with an applied AC bias current of 1 nA and without applied DC bias current. Fluctuations in the resistance become less pronounced with decreasing |*D*| and move towards smaller *n*, following the phase boundaries.

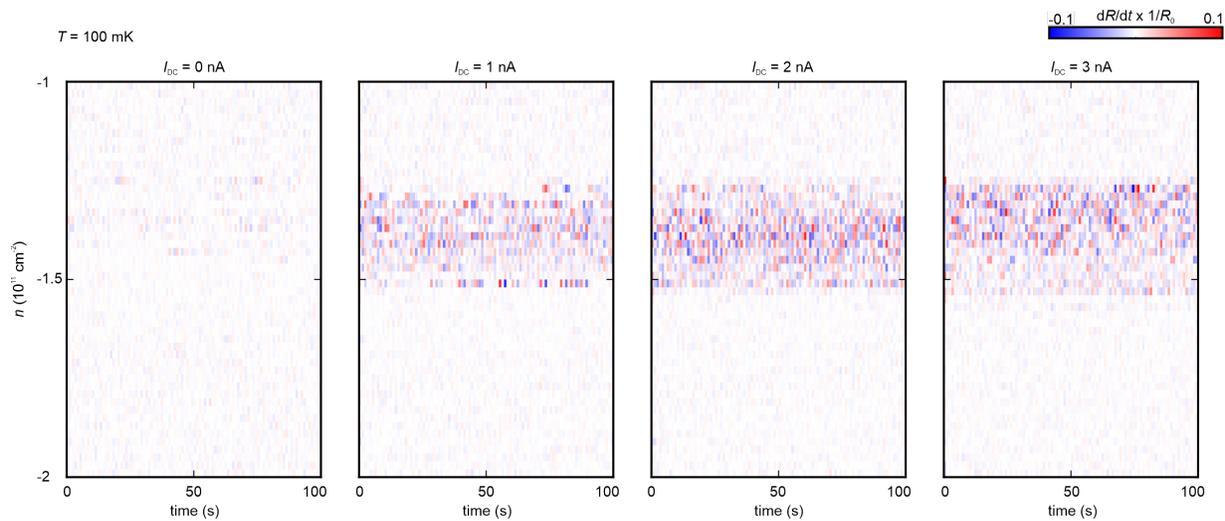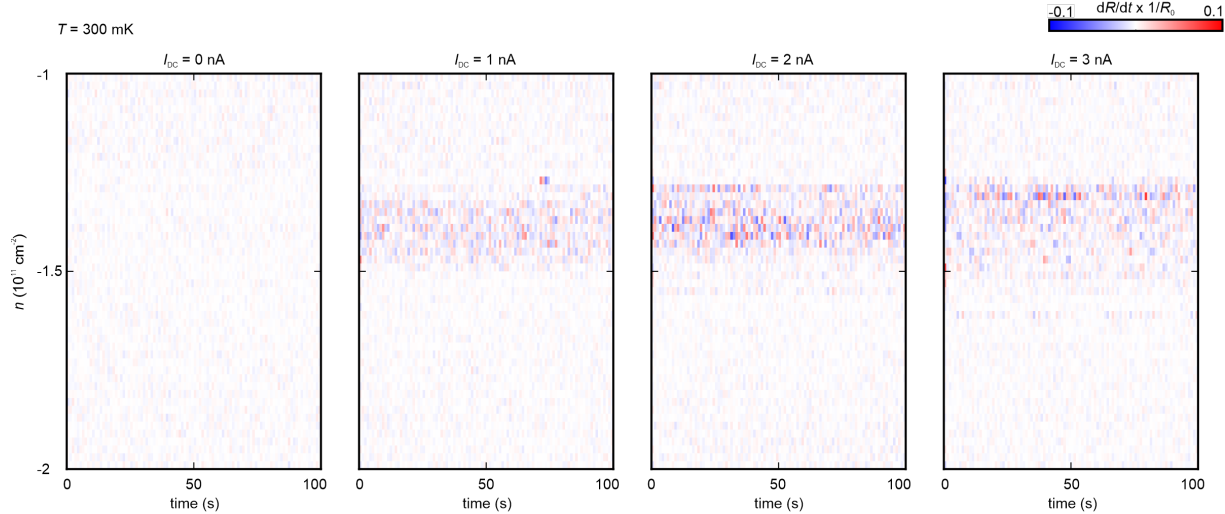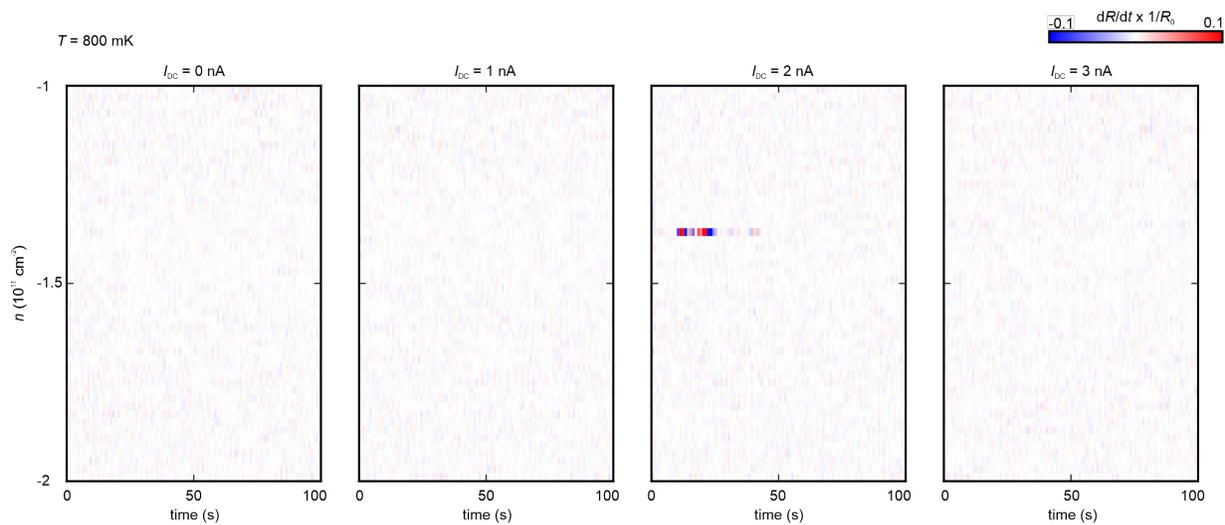

**Extended Data Fig. 2 Noise as a function of temperature.** Derivative of the normalized resistance over time dR/dt x 1/$R_0$ as a function of the charge carrier density *n* at different DC bias currents and different temperatures *T*. An AC bias current of 100 pA was applied at a frequency of 78 Hz. The electric displacement field *D* was set to -0.7 Vnm$^{-1}$. Fluctuations in the resistance become less pronounced with increasing *T*.

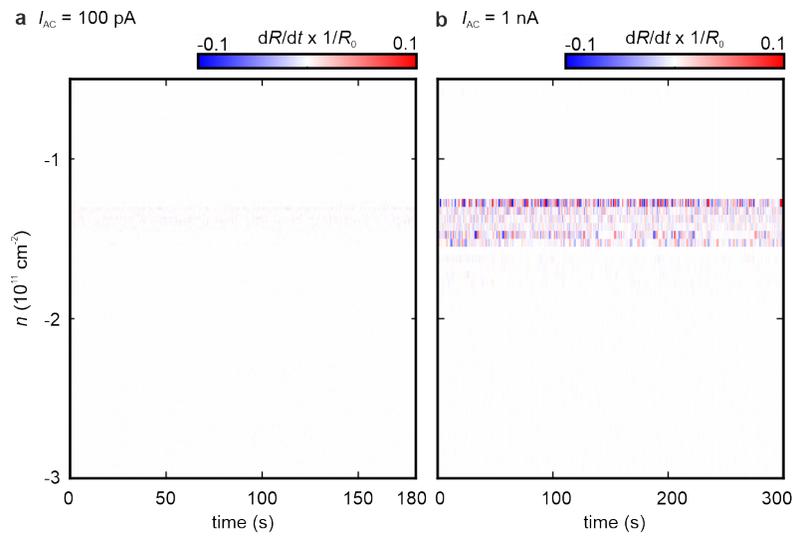

**Extended Data Fig. 3. Noise as a function of the AC currents.** Derivative of the normalized resistance over time $dR/dt \times 1/R_0$ as a function of the charge carrier density $n$ and an applied AC bias current of 100 pA **(a)** and 1 nA **(b)** at an electric displacement field of -0.7 Vnm$^{-1}$ and a frequency of 78 Hz. No DC bias current was applied.

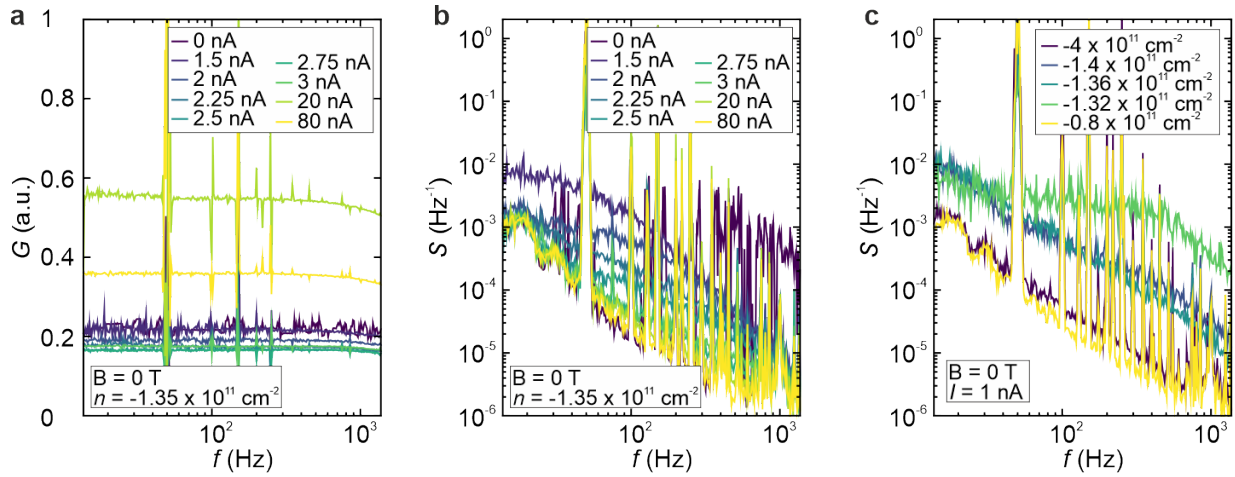

**Extended Data Fig. 4. Frequency dependence of the conductance and spectral noise density.** Conductance $G$ **(a)** and normalized spectral noise density $S$ **(b)** as a function of the applied AC frequency $f$ ($I_{AC}$ = 100 pA) measured at different DC currents $I_{DC}$ at zero magnetic field $B$, charge carrier density $n$ = -1.35 x $10^{11}$ cm$^{-2}$ and electric displacement field $D$ = -0.7 Vnm$^{-1}$. $G$ is nearly independent of $f$ for f < 1 kHz and decreases for $f$ > 1 kHz due to electrical filters integrated into the cryostat. $G$ peaks around 50 Hz and its higher harmonics arise from the electrical grid. **(b)** is identical to Fig. 4a and is shown here again for comparison purposes. **(c)** $S$ as a function of $f$ at $I_{AC}$ = 100 pA, $I_{DC}$ = 1 nA and at different $n$. S increases in the density regime of phase II ($n$ = -1.32 x $10^{11}$ cm$^{-2}$, $n$ = -1.36 x $10^{11}$ cm$^{-2}$ and $n$ = -1.4 x $10^{11}$ cm$^{-2}$) but is independent of $n$ outside of phase II.

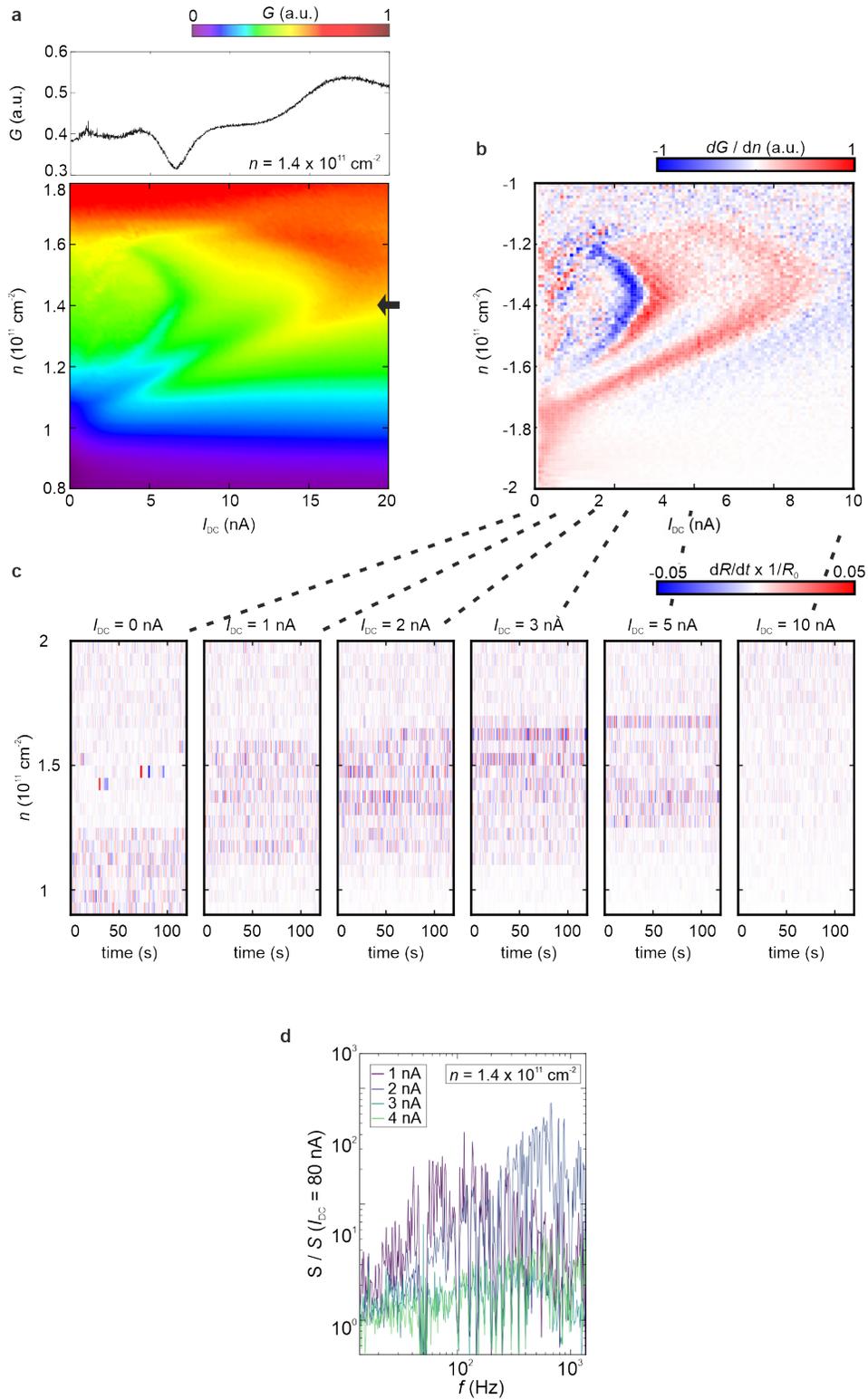

**Extended Data Fig. 5. Noise in the spin and valley-polarized insulating (svi) phase at electron-doping.**
**(a,b)** Conductance (*G*) **(a)** and derivative of the conductance (d*G*/d*I*) **(b)** as a function of the applied DC current $I_{DC}$ and the charge carrier density *n* at an electric displacement field of *D* = -0.8 Vnm$^{-1}$ and in the density regime of the spin and valley-polarized insulating (svi) phase (see Reference [22]). A linecut taken at $n = 1.4 \times 10^{11}$ cm$^{-2}$ is shown in the top. **(c)** Derivative of the normalized resistance over time $dR/dt \times 1/R_0$ as a function of time $t$ and $n$ at different $I_{DC}$. Noise, indicative of depinning of the Wigner crystal state, is present in the density regime of the svi phase at $I_{DC}$ ranging from 1 nA to 5 nA. Additionally, noise is present at $I_{DC} = 0$ and small $n$ due to the extremely high resistances within this regime that cannot be resolved by our measurement setup. **(d)** Spectral noise density $S$ normalized with respect to $S(I_{DC} = 80$ nA$)$ as a function of the applied AC frequency $f$ ($I_{AC} = 100$ pA) and measured at different $I_{DC}$ and $n = -1.4 \times 10^{11}$ cm$^{-2}$. Frequency-dependent noise bulges appear for currents ranging from 1 nA to 4 nA.

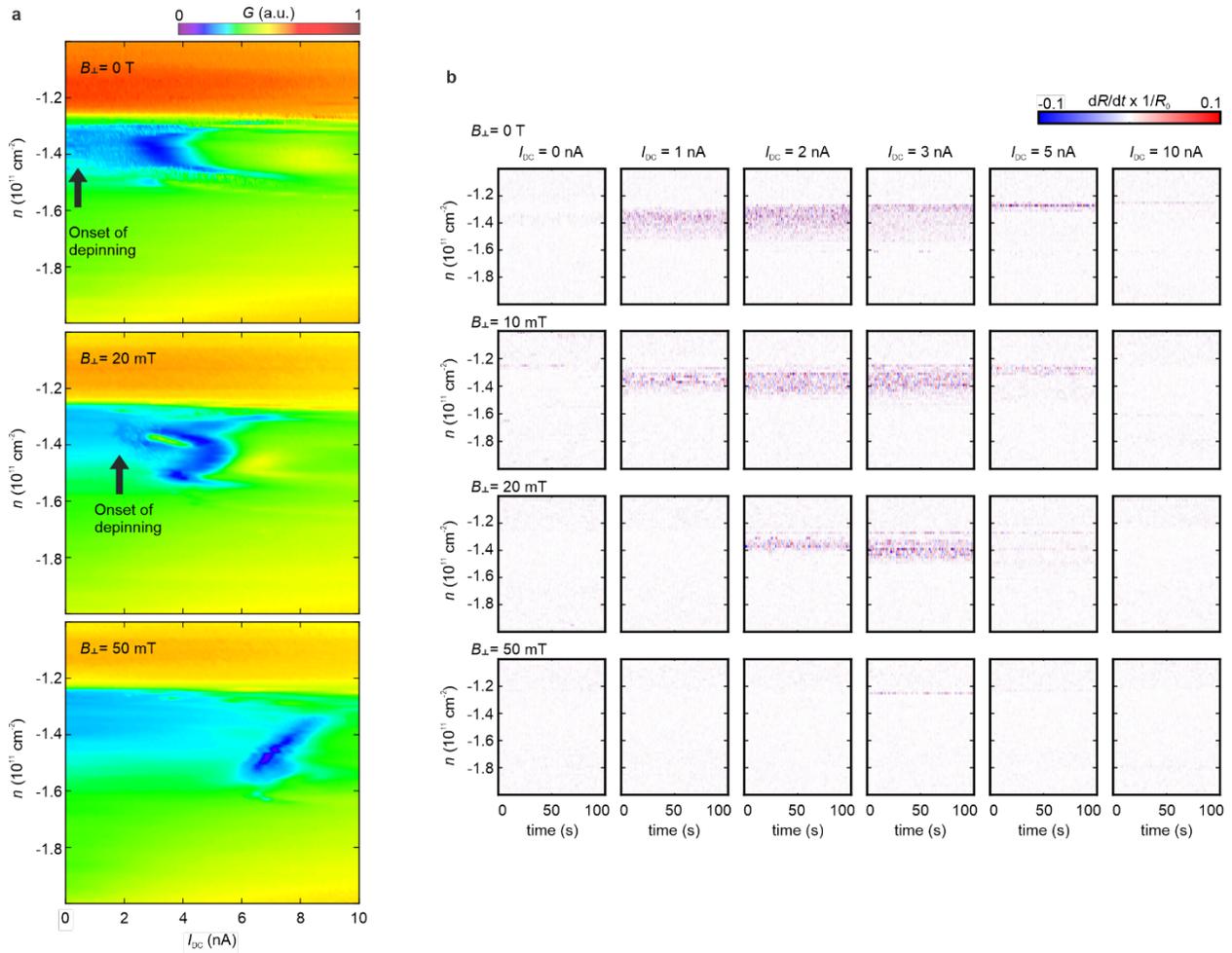

**Extended Data Fig. 6. Noise as a function of the out-of-plane magnetic field. (a)** Conductance ($G$) with as a function of the applied DC current $I_{DC}$ and the charge carrier density $n$ at different out-of-plane magnetic fields $B_\perp$. The onset of depinning moves to higher values of $I_{DC}$ with increasing $B_\perp$. **(b)** Derivative of the normalized resistance over time $dR/dt \times 1/R_0$ as a function of time $t$ and $n$ at different $I_{DC}$ and $B_\perp$. No noise, indicative of depinning of the Wigner crystal state, is present at $B_\perp$ = 50 mT.